\documentclass[preprint2]{emulateapj}

\newcommand{\myemail}{maravena@nrao.edu}

\shorttitle{Environment of MAMBO galaxies in the COSMOS field}
\shortauthors{Aravena et al.}

\begin{document}


\title{Environment of MAMBO galaxies in the COSMOS field$^{\ast}$}


\author{M. Aravena\altaffilmark{1,2,3}, F. Bertoldi\altaffilmark{2}, C. Carilli\altaffilmark{4}, E. Schinnerer\altaffilmark{5}, H. J. McCracken\altaffilmark{6}, M. Salvato\altaffilmark{7,8,9}, D. Riechers\altaffilmark{7,10}, K. Sheth\altaffilmark{1} ,V. Sm\v{o}lci\'c\altaffilmark{7}, P. Capak\altaffilmark{11} , A. M. Koekemoer\altaffilmark{12}, K. M. Menten\altaffilmark{3}}

\altaffiltext{$\ast$}{Based on observations obtained, within the COSMOS Legacy Survey, with the IRAM 30 m , NRAO-VLA, HST, CFHT, Subaru, KPNO, CTIO, and ESO Observatories. The National Radio Astronomy Observatory is a facility of the National Science Foundation (NSF), operated under cooperative agreement by Associated Universities Inc.}
\altaffiltext{1}{National Radio Astronomy Observatory. 520 Edgemont Road, Charlottesville VA 22903, USA. \myemail}
\altaffiltext{2}{Argelander Institut f\"ur Astronomie. Auf dem H\"ugel 71, 53121 Bonn, Germany.}
\altaffiltext{3}{Max-Planck Institut f\"ur Radioastronomie. Auf dem H\"ugel 69, 53121 Bonn, Germany.}
\altaffiltext{4}{National Radio Astronomy Observatory. P.O. Box O, Socorro, NM 87801, USA}
\altaffiltext{5}{Max-Planck-Institut für Astronomie. K\"onigstuhl 17, D-69117 Heidelberg, Germany}
\altaffiltext{6}{Institut d'Astrophysique de Paris. 98bis boulevard Arago, 75014 Paris, France}
\altaffiltext{7}{California Institute of Technology. 1200 East California Boulevard, Pasadena, CA 91125, USA}
\altaffiltext{8}{Max-Planck Institut f\"ur Plasmaphysik. Boltzmannstrasse 2, D-85748, Garching, Germany}
\altaffiltext{9}{Excellence Cluster Universe, Boltzmannstrasse 2. D-85748, Garching, Germany}
\altaffiltext{10}{Hubble fellow}
\altaffiltext{11}{Spitzer Science Center, Caltech, Pasadena, CA 91125, USA}
\altaffiltext{12}{Space Telescope Science Institute. 3700 San Martin Drive, Baltimore, MD 21218, USA}


\begin{abstract}
Submillimeter galaxies (SMG) represent a dust-obscured high-redshift
population undergoing massive star formation activity. Their
properties and space density have suggested that they may evolve into
spheroidal galaxies residing in galaxy clusters. In this paper, we
report the discovery of compact ($\sim10-20\arcsec$) galaxy
overdensities centered at the position of three SMGs detected with the
Max-Planck Millimeter Bolometer camera (MAMBO) in the COSMOS
field. These associations are statistically significant. The
photometric redshifts of galaxies in these structures are consistent
with their associated SMGs; all of them are between $z=1.4-2.5$,
implying projected physical sizes of $\sim170$ kpc for the
overdensities. Our results suggest that about 30\% of the
radio-identified bright SMGs in that redshift range form in galaxy
density peaks in the crucial epoch when most stars formed.
\end{abstract}





\section{Introduction}

Submillimeter galaxies (SMG) are dust-obscured starburst galaxies at
high-redshift \citep{Smail1997, Hughes1998, Barger1998}. Their large dynamical
\citep{Greve2005, Tacconi2006, Tacconi2008} and stellar masses
\citep{Borys2005, Dye2008}, as well as their number densities and
clustering properties \citep{Scott2002, Blain2004, Viero2009}, suggest
they could be the progenitors of present-day luminous ellipticals
\citep{Lilly1999, Swinbank2006}. Some SMGs are known to be associated
with galaxy clusters at high redshifts \citep[e.g.][]{Webb2005} and to
be located in extended overdensities of LBGs and radio galaxy fields
\citep{Ivison2000, Smail2003, Stevens2003, Chapman2008, Daddi2009,
  Tamura2009}.

If SMGs are progenitors of massive clustered spheroids, they would
likely show signs of clustering in the epoch when these galaxies have
their peak in activity and luminosity ($z\sim 1 - 3$), similar to what
is observed for powerful AGN \citep{Miley2008}. Attempts to measure
the clustering of SMGs indicate that they are associated with massive
dark matter halos and possibly trace the largest scale structures at
high redshifts \citep{Blain2004, Viero2009, Weiss2009}. However,
current submillimeter blank-field surveys either cover small areas in
the sky, yielding a few tens of sources within contiguous fields
\citep{Coppin2006, Bertoldi2007, Scott2008, Perera2008,
  Austermann2009}, or are limited by poor angular resolution albeit
covering large regions \citep{Devlin2009}. Hence, good quantitative
studies of the small to large scale clustering of the SMG population
are not feasible until large surveys comprising a few square degrees
on the sky under good resolution ($\lesssim20\arcsec$) can be made.

Studies of the environment of SMGs are possible when utilizing the
rich complementary data available for current (sub)millimeter
fields. Whether SMGs are embedded in regions with an enhanced number
of optical/near-IR detected high-redshift galaxies has yet not been
quantified. In this paper, we investigate to what extent SMGs are
located in clustered fields. For this, we make use of deep optical and
near-IR imaging data in the central part of the COSMOS field to
measure the density of high-redshift $BzK$ galaxies in the
surroundings of SMGs. This allows us to study the relation of
blank-field detected SMGs with the most prominent galaxy density peaks
at high-redshift. Hereafter, we assume a cosmology with $H_0=70$ km
s$^{-1}$ Mpc$^{-1}$, $\Omega_{\Lambda}=0.7$ and
$\Omega_\mathrm{M}=0.3$ and use all magnitudes in the AB system.


\section{Observations}
\label{sect:observations}
\subsection{COSMOS photometric data}

\begin{figure}[!t]
 \centering
\includegraphics[scale=0.52]{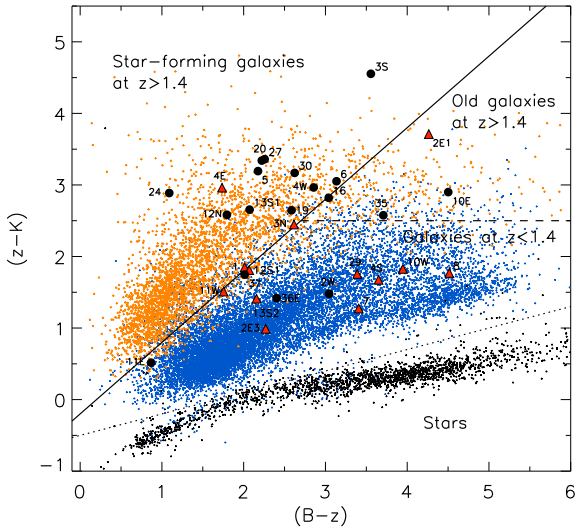}
\caption[]{$BzK$ diagram for MAMBO sources in the COSMOS
  field. Reliable $K$-band selected identifications are represented by
  filled black circles and ambiguous or unreliable identifications are
  shown as filled red triangles. The BzK loci for galaxies are shown
  as small crosses. Their color coding (orange, blue and black) is
  based on the photometric redshifts ($z>1.4$, $z<1.4$ and stars,
  respectively). The solid line separates star-forming galaxies at
  $z>1.4$, while the dashed line separates old passive evolving
  galaxies at $z>1.4$. The dotted line separates stars (black) from a
  mixed population of galaxies at $z<1.4$.  \label{fig:bzk}}
\end{figure}

The COSMOS survey \citep{Scoville2007a} covers a sufficiently large area
,$1.4\degr\times1.4\degr$, at appropriate depth over nearly the entire
electromagnetic spectrum to provide a comprehensive view of galaxy
formation and large scale structure.

The COSMOS $i^+$-band selected photometric catalog includes a total of
30 narrow, intermediate and broad band filters covering from UV to
mid-IR wavelengths that allowed the computation of accurate
photometric redshifts down to $i_\mathrm{AB}^{+}=26.5$
\citep[see][]{Ilbert2008,Salvato2009}.

A new deep $K$-band survey covering the entire COSMOS field was
recently carried out by \citet{McCracken2009}. The $K$-band image
reaches seeing values of $\sim0.7$\arcsec \ with variations of less
that $\sim20\%$ across the entire COSMOS field. The $K$-band selected
catalog includes the $B^+$, $z^+$ and $i^+$ bands, reaching a
completeness limit $K_\mathrm{AB}\approx23.0$ and a 1$\sigma$ limit
in the 2\arcsec \ diameter aperture of 25.4 \citep[for
  details see][]{McCracken2009}. About 85\% of the sources detected in
this $K$-band image down to $K=23.0$ have a photometric redshift
estimate based on the $i^+$-band photometric redshift catalog.

The optical/IR imaging was supplemented with deep Very Large Array (VLA) 1.4 GHz radio imaging \citep{Schinnerer2004, Schinnerer2007, Bondi2008} for the full COSMOS area to an average rms level of 10 $\mu$Jy at a resolution of $1.5\arcsec$.

\subsection{MAMBO 1.2 mm observations}
An effective area of $\sim22\arcmin\times22\arcmin$ of the COSMOS
field was mapped at 1.2 mm by \citet{Bertoldi2007} using the
Max-Planck millimeter bolometer camera (MAMBO) at the Institut de
Radioastronomie Millim\'etrique (IRAM) 30 m telescope. The COSMOS
MAMBO (COSBO) survey is centered at (R.A., Dec.)
$=(10^{\mathrm{h}}\ 00^{\mathrm{m}}\ 30^{\mathrm{s}},\ 02\degr\ 12\arcmin\ 00\arcsec)$
and reaches a noise level of 1.0 mJy per $11\arcsec$ beam.

Fifteen millimeter sources were detected with a significance
$>4\sigma$. Eleven of them were found to have a significant 1.4 GHz
VLA counterpart. Additional 12 lower significance sources
($3.5-4.0\sigma$) were selected based on their association with radio
sources within $5\arcsec$ from the millimeter position
\citep{Bertoldi2007}. Two more MAMBO sources with S/N$=3-4\sigma$
were selected based on their match with significant 1.1 mm Bolocam
(Aguirre et al., in prep.) and radio sources.

\begin{figure}[!t]
 \centering
 \includegraphics[scale=0.45]{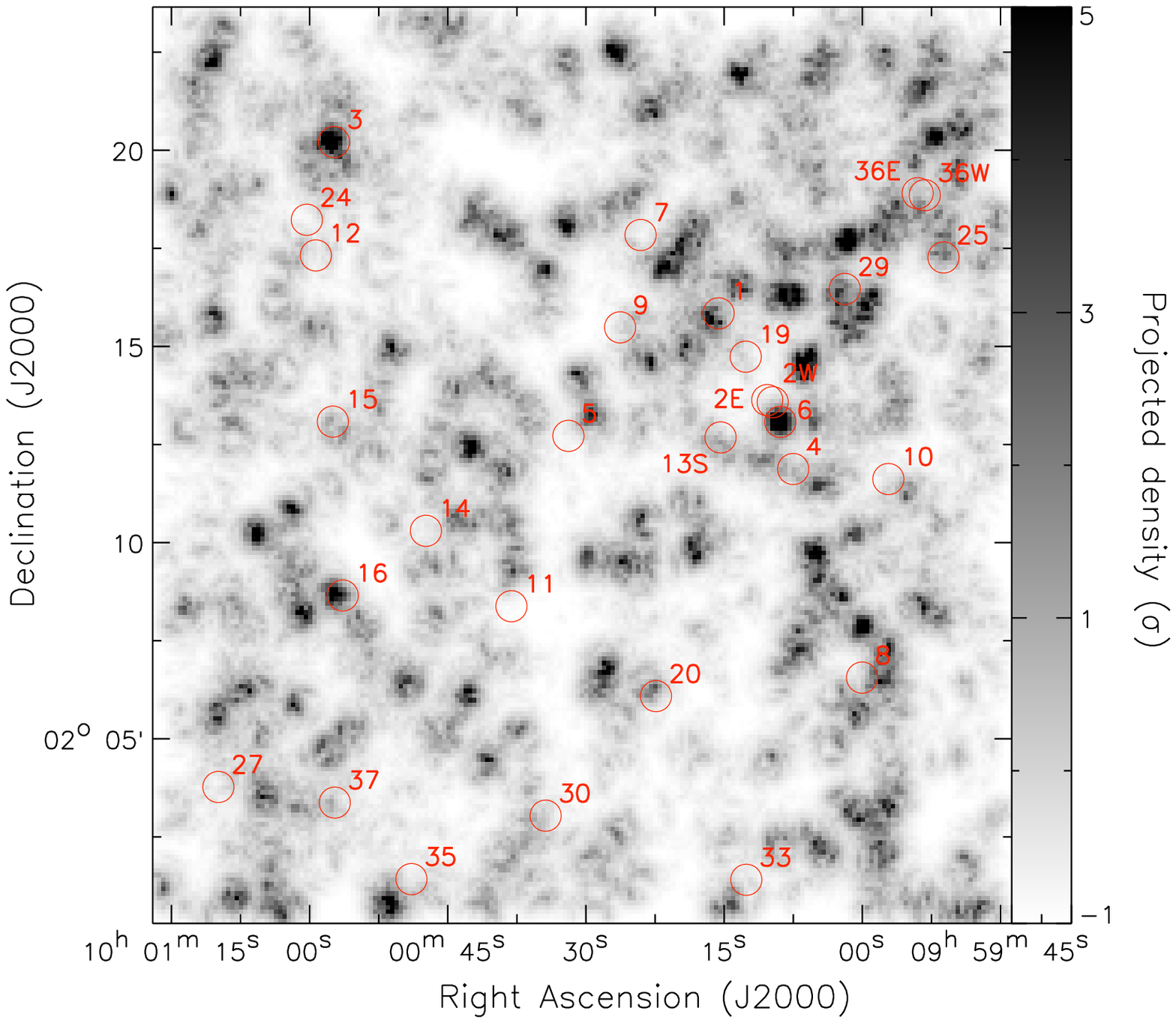}
 \includegraphics[scale=0.45]{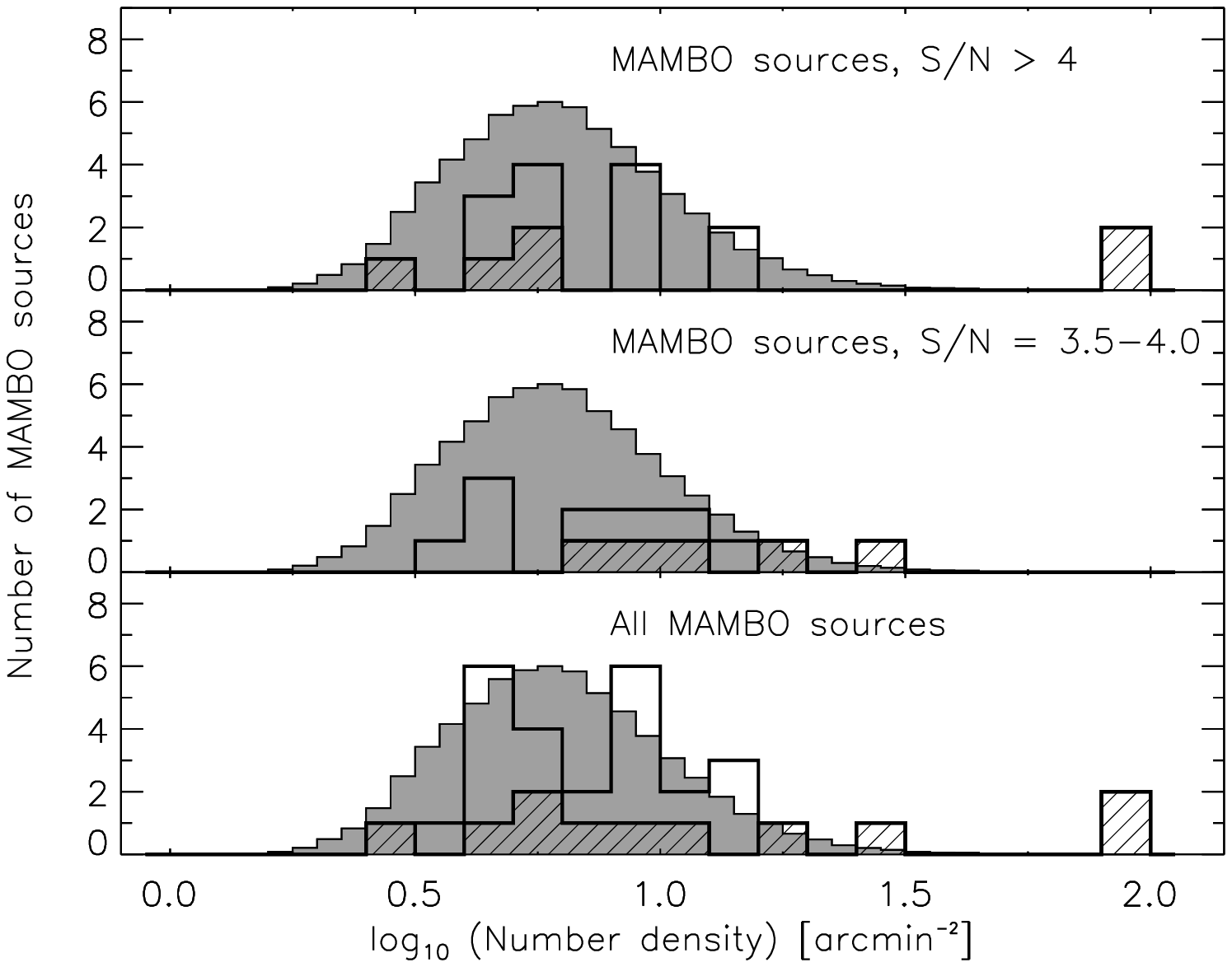}
 \caption{\textit{Top}: Projected number density of high-redshift
   $BzK$ galaxies in the COSBO field ($K<23$). The gray-scale
   represents the density map given in terms of the standard deviation
   $\sigma$ with respect to the average background level. Red circles
   mark the position of the MAMBO galaxies. ID numbers are the same as
   in \citet{Bertoldi2007}. \textit{Bottom}: Distribution of densities
   of high-redshift $BzK$ galaxies at the position of MAMBO sources
   (solid) and MAMBO sources with photometric redshifts in the range
   $1.4-2.5$ (solid hatched) compared to the distribution of densities
   of BzK galaxies obtained from the $200\times200$ grid points in the
   COSBO density map (shaded). \label{fig:density_map_highz}}
\end{figure}

\section{Environment of submillimeter galaxies}

\subsection{$BzK$ selection}
\label{sect:BzKsel}

The $BzK$ color-color criterion \citep{Daddi2004} provides an
efficient way to select galaxies in the crucial epoch when
star-formation and SMG activity peaked ($z=1-3$).

Figure \ref{fig:bzk} shows the $BzK$ color-color diagram for the
$K$-band selected counterparts to the MAMBO sources and for objects
with $K<23$ in the COSBO field. Small corrections account for the
difference between the VLT $B$-band used in the original $BzK$
criterion and our $B_J$ band \citep{McCracken2009}. According to this,
sources with $BzK \equiv (z-K)-(B-z) > -0.2$ are star-forming galaxies
at redshift $>1.4$ (s$BzK$), while objects with $BzK<-0.2$ but
$(z-K)>2.5$ are old passively evolving galaxies at redshift $>1.4$
(p$BzK$). Objects with $BzK<-0.2$ and $(z-K)<2.5$ correspond to a
mixture of old and star-forming galaxies at redshift $<1.4$
(n$BzK$). The restriction $(z-K) < 0.3(B-z)-0.5$ allows the separation
of stars.

To study the environment of the MAMBO galaxies, we used the
$BzK$ criteria to create a $K$-band selected galaxy sample containing
a mixture of passive and star-forming galaxies at high-redshift. We constrained our sample to include objects
in the magnitude range $K=17-23$ and used the $BzK$ criterion to
reject stars.

We note that while the $BzK$ criterion is very efficient selecting
galaxies at $z>1.4$, with $\sim90\%$ completeness down to $K=23$
\citep[based on spectroscopic measurements;][]{Barger2008}, it still
suffers a great degree of contamination ($\sim36\%$), mostly from
sources at $z=1.0-1.4$ \citep[$\sim30\%$;][]{Barger2008}. Only a small
percentage ($\sim10\%$) appears to lie at $z<1$ and $z>3$.

\begin{table*}[!t]
\centering
\caption{Associations between high-redshift $BzK$ galaxy overdensities with millimeter sources\label{table:significance}}
\begin{tabular}{llllllllll}
\hline\hline
 ID & R.A.$^a$ & Dec.$^a$ & $N_{30\arcsec}^b$ & $S/N^c$& $d^d$& $P_{30\arcsec}^e$& $P_d^f$ & $z_\mathrm{median}^{g}$ & $z_\mathrm{SMG}^{h}$ \\  
          & \multicolumn{2}{c}{(J2000)} & & & (\arcsec) &  $\times10^{-2}$ & $\times10^{-2}$ &  & \\
\hline
COSBO-1  &  150.0679 & 2.26209 & 9 &  4.6 & 13.8 & 18.5 &  4.7           & 2.0 & 1.2   \\
COSBO-3  &  150.2379 & 2.33649 & 10 & 27.8 & 2.8 & 3.9  &  0.05          & 2.3 & 2.3  \\
COSBO-6  &  150.0357 & 2.21871 & 11 & 22.9 & 4.6 & 12.3 &  0.2           & 1.6 & 1.9  \\
COSBO-16 &  150.2360 & 2.14549 & 12 &  7.2 & 6.5 & 12.7 &  0.7           & 1.5 & 0.5  \\
\hline
\end{tabular}\\
\begin{flushleft}
\begin{footnotesize}
\noindent $^a$ Position of the density peak. $^b$ Number of
high-redshift $BzK$ galaxies within 30\arcsec. $^c$ Peak
signal-to-noise ($S/N$) of the density peak. $^d$ Distance from the
COSBO source to the density peak. $^{e \& f}$ Probability that a
significant overdensity lies by chance within a distance of 30\arcsec \ and of
$d\arcsec$ from the MAMBO source.$^g$ Median photometric redshift for $BzK$ galaxies within 30\arcsec. $^h$ Photometric redshift for the SMG from \citet{Bertoldi2007}. 
\end{footnotesize}
\end{flushleft}
\end{table*}

\begin{figure*}[!t]
 \centering
\includegraphics[scale=0.95]{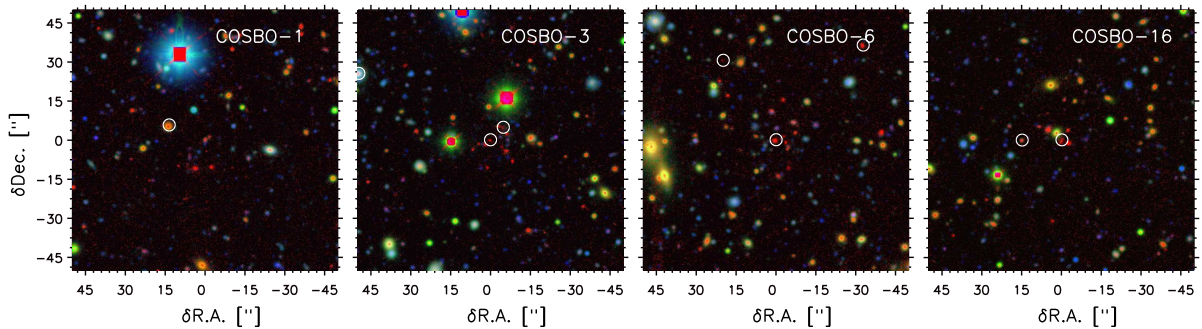}
\caption{$BzK$ color images of MAMBO galaxy fields that are related to strong overdensities of high-redshift galaxies. The images are centered at the MAMBO source position. White circles show the position of significant VLA 1.4 GHz sources.\label{fig:master}}
\end{figure*}

\subsection{Projected density distribution}
\label{sect:overdensities2}

We constructed number density maps of high-redshift $BzK$ galaxies in
the COSBO field. For this, we created a grid of $200\times200$
positions equally spaced by 7\arcsec \ centered at the COSBO field
position, and computed the density from the distance to the 7th
nearest neighbour $BzK$ galaxy, $d_7$, to each grid point. The density
is thereby computed as $ n_{7}=7/\pi (d_{7})^{2} $.

This procedure is similar to the one introduced by
\citet{Dressler1980}. The choice of $7$ is a compromise between
accounting for structures on small scales (groups of $\gtrsim4$
galaxies around SMGs) and good statistics for each density
value. Similar approaches have been applied to known rich galaxy
clusters \citep[e.g.][]{Guzzo2007}, but using photometric or
spectroscopic information to compute the number density in redshift
slices.

Figure \ref{fig:density_map_highz} shows the projected number density
map of high-redshift $BzK$ galaxies in the COSBO
field. The average and standard deviation values were computed by
using all grid points from our density map
(Fig. \ref{fig:density_map_highz}). The average density thus computed
is $7.0\pm4.6$ arcmin$^{-2}$ ($K<23.0$), where the quoted uncertainty
corresponds to the actual cosmic variance. Down to $K=21.8$, we find
an average density of $1.5\pm1.0$ arcmin$^{-2}$, which compares well
with the number counts of high-redshift $BzK$ galaxies in other
surveys at this depth \citep{Kong2006}.

We find that four MAMBO galaxies (COSBO 1, 3, 6 and 16) are embedded
in significant overdensities ($>4\sigma$) of high-redshift $BzK$
galaxies. All the galaxies in these overdensities are star-forming
rather than passive evolving galaxies. Three out of these four SMGs
were detected with S/N$>4$ in the MAMBO map \citep{Bertoldi2007}. The
overdensities of galaxies around MAMBO sources can also be seen
directly on the optical and IR images (Fig. \ref{fig:master}). Here,
red galaxies are easily distinguished, showing an excess toward the
MAMBO source position (image center). The typical radial extent of
these overdensities is $\sim5-10$\arcsec \ ($\sim20\arcsec$ in
diameter). At a redshift of $\sim2$, this implies structures on scales
of $\sim170$ kpc, similar to what has been found for QSO and radio
galaxy fields \citep{Hall1998, Best2000}.

\subsection{Probability of chance association}
\label{sect:probability}
To estimate the probability $P$ that an overdensity is found by chance
within a distance $d$ from a MAMBO galaxy, we performed Monte Carlo
simulations of significant density peaks drawn by the underlying
distribution of overdensities in the COSBO field.

To identify all the peaks in our density map, we used the IDL version
of the DAOPHOT task FIND. This routine finds the positive
perturbations in the density map, and uses marginal Gaussian fits to
locate the centroid and amplitude of the density peak. This Gaussian
approximation may not be valid for fragmented density structure,
however it is reliable to detect the most significant overdensities
which are typically non-fragmented. 

Based on this procedure, we found 45 detections with S/N$>4$, nine of
them with S/N$\gtrsim10$.  Using their observed spatial distribution,
we generated 45 peaks in each of the 10000 samplings, and thereby
computed $P$ as the fraction of SMG-overdensity associations in our
simulations (Table \ref{table:significance}). In the cases of COSBO 3,
6 and 16, the probabilities of chance association are negligible.

\subsection{Photometric redshifts and comments on individual associations}
\label{sect:photzenv}

To measure the clustering of the star-forming high-redshift galaxies
associated with SMGs in redshift space, we used the COSMOS $i^+$-band
selected photometric redshift catalog \citep{Ilbert2008}. The accuracy
of these photometric redshifts for faint galaxies ($i^+\gtrsim25.5$)
is $\sigma\sim0.2$.

\begin{figure}[!t]
 \centering
\includegraphics[scale=0.5]{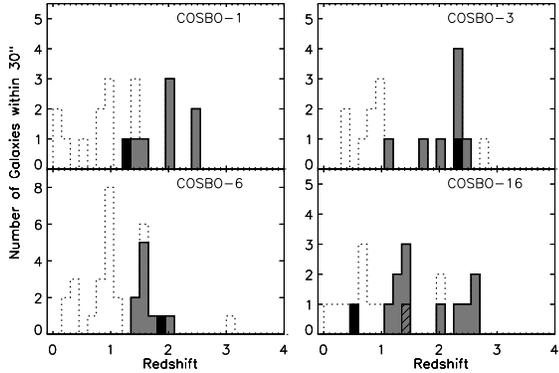}
\caption{Redshift distribution for galaxies located within 30\arcsec
  \ from the density peak of high-redshift galaxies near MAMBO
  sources. The dotted histogram shows the photometric redshifts for
  our $K$-band selected galaxies, the dark-gray histogram shows the
  redshifts for the $K$-band selected high-redshift $BzK$ galaxies and
  the black single object shows the redshift of the most likely
  counterpart to the MAMBO source as given by
  \citet{Bertoldi2007}. For COSBO-16, the hashed entry represents the
  secondary solution from the photometric redshift
  computation. \label{fig:photo_z}}
\end{figure}

Figure \ref{fig:photo_z} shows the redshift distribution of $K$-band
selected galaxies that lie close to the high-redshift $BzK$ density
peaks associated with SMGs. Since the $i^+$-band catalog is too
shallow to include the obscured optical emission from the MAMBO
galaxies, we used the previous photometric redshift estimates from
\citet{Bertoldi2007}.

For COSBO-1, a recent Submillimetre Array (SMA) detection (which will
be published elsewhere) indicates that the millimeter emission is
produced by a radio and optically undetected galaxy at $z>3.5$
\citep[as also implied by its radio-to-millimeter spectral
  index;][]{Bertoldi2007}. The counterpart to this MAMBO source
selected by \citet{Bertoldi2007}, a relatively bright IRAC/optical
source with a photometric redshift of $\approx1.2$, is related to the
galaxy group at this redshift. However, it is very probably not
responsible for the millimeter emission. The association between this
source and the overdensity is unlikely, $P_{d}\sim0.05$. Hence, we
discard this one as a real association between a SMG and a galaxy
group at $z\sim1.5$. In cases when the group is at low-redshift,
gravitational lensing of a far-away SMG is a possibility.

For COSBO-3, the redshift distribution of the galaxies in its close
neighbourhood is consistent with most of them being at $z\sim2.2-2.4$
(Fig. \ref{fig:photo_z}). Two radio sources can be identified within
10\arcsec \ from the MAMBO source. The most likely radio/IR/optical
counterpart (COSBO-3S) has a photometric redshift of $2.3$
\citep{Bertoldi2007}. Based on the $i^+$-band selected catalog, we
find a redshift of 2.4 for this source, consistent with the likely
redshift of the galaxy group. Recent CARMA observations indicate the
MAMBO emission is produced by at least two sources (Sm\v{o}lci\'c et
al., in prep.).

Most galaxies in the group around COSBO-6 lie at
$z\sim1.2-1.8$. \citet{Bertoldi2007} estimated a photometric redshift
of $1.9$ for the radio/optical identified counterpart to the
millimeter emission. Although the redshift for the likely counterpart
is slightly larger than the one implied by the redshift distribution
of the surrounding high-redshift $BzK$ galaxies, it agrees within
$\Delta z = 0.2$.

The galaxies surrounding COSBO-16 have a photometric redshift of
$\sim1.4$. The photometric redshift derived for the likely
radio/optical counterpart from the $i^+$-band selected catalog is
$\approx0.5$ \citep{Bertoldi2007}, however the secondary solution
(second minima in the $\chi^2$ distribution) implies $z\sim1.4$. The
photometric redshift listed in the COSMOS catalog is, however, 2.55
which agrees better with that implied by the radio to millimeter
spectral index \citep{Bertoldi2007}. Because of the somewhat ambiguous
redshift derivation for the millimeter source, it is difficult to
relate it to the redshift peak in the surrounding galaxies, although
we may slightly favor to use $z=1.4$ as the most likely case for both.

\section{Summary and Discussion}
\label{sect:discussion}

We find that significant overdensities of star-forming high-redshift
galaxies are related to three MAMBO galaxies detected in the COSMOS
field.  These groups are compact in size, and the peaks of their
redshift distributions are compatible with the redshift estimated for
the associated MAMBO galaxies.

\subsection{SMGs in dense environments}

If SMGs are related to the formation of structures at high-redshift,
we would expect that most SMGs are located in regions with enhanced
galaxy densities. However, only a few SMGs in our sample can be
associated with strong galaxy overdensities. This could be partly
attributed to a selection effect, since we are comparing the overall
population of SMGs at various redshifts with $K$-band selected
galaxies in the range $z\sim1.4-2.5$. According to
\citet{Chapman2005}, $\sim55\%$ of the radio-identified, bright SMG
population lies in this redshift range. From our sample of fifteen
MAMBO sources detected with a significance $>4\sigma$, eleven were
identified to have a radio counterpart. Following Chapman et al., we
estimate that about six of these eleven MAMBO sources should be at
$z=1.4-2.5$, while using the photometric redshifts reported by
\citet{Bertoldi2007}, seven appear to lie in this redshift range. This
implies that $\sim$30\% of the radio-identified, significant MAMBO
sources at $z=1.4-2.5$ are associated with substancial overdensities
of massive galaxies at these redshifts.

Note that our study is biased in that we are selecting massive
galaxies at $z=1.4-2.5$. Our $K<23$ limit roughly translates into
stellar masses $\gtrsim (2-4)\times10^{10}\ M_{\sun}$
\citep{Daddi2004}, and therefore we miss less massive galaxies that
could be associated with SMGs at these redshifts. However, the fact
remains that only a fraction of the SMGs are related to groups of
massive galaxies in the crucial epoch of galaxy assembly.

We compared the distribution of densities at the position of
$>4\sigma$ MAMBO source detections with the distribution of densities
of BzK galaxies in the field (Fig. \ref{fig:density_map_highz}). A
Kolmogorov-Smirnov (KS) test does not reject the null hypothesis that
both samples follow the same distribution at a 46\% significance
level. Restricting the $>4\sigma$ MAMBO source sample to sources at
$z=1.4-2.5$, the KS test does not reject the null hypothesis at a 43\%
level. This is somewhat inconsistent with the fact that two of the
brightest MAMBO galaxies are related to some of the strongest
overdensities in the field, however it may merely reflect the low
number of SMGs that we used to compute the KS statistic. For the
$3.5-4.0\sigma$ MAMBO source sample, the KS-test does not rule out the
null hypothesis at a 23\% level, whereas if we limit this sample to
galaxies with $z=1.4-2.5$, the KS test gives a 3.4\% significance
level. This strongly suggests that the densities at the position of
the fainter MAMBO sources follow a different distribution from that of
BzK galaxies in the field. For both samples combined (all MAMBO
sources), the KS test gives only a 9.6\% probability that they follow
the same distribution of densities of BzK galaxies in the field, which
is similar to the value that we obtain if we restrict the sample to
$z=1.4-2.5$, 12.6\%, again suggesting that the distributions are
different.

Overall, our results show that only a fraction (30\%) of MAMBO sources
at $z=1.4-2.5$ is located in strongly overdense regions. This suggests
that only some SMGs are linked to the formation of structures
at high-redshift. Although we find a hint that some SMGs could be located in
environments denser than that of the general population of galaxies at
$z\sim2$, it is not possible to discern whether this is a real trend
or is due to cosmic variance as our analysis is based only on a
handful of SMGs.

The presence of bright SMGs in galaxy overdensities is possibly
related to the fact that the density peaks of high-redshift $BzK$
galaxies associated with MAMBO galaxies (e.g. for COSBO-3 and 6) are
the strongest in the whole COSBO area. Denser groups of star-forming
$BzK$ galaxies are more likely to produce major mergers between
galaxies and thus are prone to trigger violent star-formation
activity. The merger of two gas-rich $BzK$ galaxies close to the
center of a galaxy group could induce a starburst that would be seen
as a bright SMG. Star-forming $BzK$ galaxies have large reservoirs of
molecular gas \citep[$\sim10^{11} M_{\sun}$;][]{Daddi2008}, and could
easily sustain the typical star-formation rates observed in SMGs
($\sim1000\ M_{\sun}$ yr$^{-1}$) for $\lesssim100$ Myr. The densest
galaxy groups are thus ideal for the formation of bright submillimeter
activity.

\subsection{Comparison with other studies}

Similar studies relating the distribution of SMGs with the large scale
structures traced by optically selected galaxies at high-redshift have
recently been done. In particular, \citet{Tamura2009} found a strong
overdensity of SMGs toward a massive protocluster of Lyman-$\alpha$
emitters at $z=3.1$, reflecting a strong link between SMGs and
the large scale structure. Nevertheless, this study is based on a
proto-cluster field where we know that strong clustering is taking place.

Studies of the galaxy-galaxy angular correlation function in
blank-fields indicate that SMGs and IR luminous galaxies at $z\sim2$
are clustered on typical angular scales of $15-25\arcsec$, being
related to massive dark matter halos
\citep[$\sim10^{13}\ M_{\sun}$;][]{Blain2004, Greve2004,Scott2006,
  Farrah2006, Viero2009, Weiss2009}. Our results are consistent with
these results in that strong clustering between SMGs and BzK galaxies
occur on similar angular scales. We note, however, that studies purely
based on the angular two-point correlation function are only able to
measure the average clustering properties of SMGs. They miss the
important fact that not all the SMGs are located in clustered
environments, as we find in this paper, and therefore only a few of
them will significantly contribute to the clustering signal of the
angular correlation function in small scales.

\acknowledgments M. Aravena was partly supported for this research
through a stipend from the International Max-Planck Research School
(IMPRS) for Radio and Infrared Astronomy at the Universities of Bonn
and Cologne. D. Riechers acknowledges support from NASA through Hubble
Fellowship grant HST-HF-01212.01A awarded by the STScI, operated by
AURA, under contract NAS 5-26555.


\clearpage

\end{document}